# Origin and Dynamical Evolution of Neptune Trojans – I: Formation and Planetary Migration

Patryk Sofia Lykawka,[1*] Jonathan Horner,[2] Barrie W. Jones [2] and Tadashi Mukai[1]

[1] Dept. of Earth and Planetary Sciences, Kobe University, 1-1 rokkodai-cho, nada-ku, Kobe 657-8501, Japan

[2] Dept. of Physics and Astronomy, The Open University, Walton Hall, Milton Keynes, MK7 6AA, UK

---

[*] Present address: International Center for Human Sciences (Planetary Sciences), Kinki University, 3-4-1 Kowakae, Higashiosaka-shi, Osaka-fu, 577-8502, Japan. E-mail address: patryksan@gmail.com

**ABSTRACT**

We present the results of detailed dynamical simulations of the effect of the migration of the four giant planets on both the transport of pre-formed Neptune Trojans, and the capture of new Trojans from a trans-Neptunian disk. The cloud of pre-formed Trojans consisted of thousands of massless particles placed on dynamically cold orbits around Neptune's L4 and L5 Lagrange points, while the trans-Neptunian disk contained tens of thousands of such particles spread on dynamically cold orbits between the initial and final locations of Neptune. Through comparison of the results with previous work on the known Neptunian Trojans, we find that scenarios involving the slow migration of Neptune over a large distance (50 Myr to migrate from 18.1 AU to its current location, using an exponential-folding time of $\tau = 10$ Myr) provide the best match to the properties of the known Trojans. Scenarios with faster migration (5 Myr, with $\tau = 1$ Myr), and those in which Neptune migrates from 23.1 AU to its current location, fail to adequately reproduce the current day Trojan population. Scenarios which avoid disruptive perturbation events between Uranus and Neptune fail to yield any significant excitation of pre-formed Trojans (transported with efficiencies between 30 and 98% whilst maintaining the dynamically cold nature of these objects – $e < 0.1$, $i < 5°$). Conversely, scenarios with periods of strong Uranus-Neptune perturbation lead to the almost complete loss of such pre-formed objects. In these cases, a small fraction (~0.15%) of these escaped objects are later recaptured as Trojans prior to the end of migration, with a wide range of eccentricities (<0.35) and inclinations (<40°). In all scenarios (including those with such disruptive interaction between Uranus and Neptune) the capture of objects from the trans-Neptunian disk (through which Neptune migrates) is achieved with efficiencies between ~0.1 and ~1%. The captured Trojans display a wide range of inclinations (<40° for slow migration, and <20° for rapid migration) and eccentricities (<0.35), and we conclude that, given the vast amount of material which undoubtedly formed beyond the orbit of Neptune, such captured objects may be sufficient to explain the entire Neptune Trojan population.

# 1 INTRODUCTION

Objects which orbit the Sun either 60º ahead or behind a planet in its orbit are known as "Trojans". These objects, moving within the 1:1 mean-motion resonance (hereafter MMR) of that planet, librate around the locations of its L4 and L5 Lagrange points, and can be dynamically stable on timescales of millions, or even billions of years (Holman & Wisdom 1993; Murray & Dermott 1999; Nesvorny & Dones 2002). There are two distinct types of Trojan behaviour – tadpole orbits, in which the object in question librates around just one of the Lagrange points, and horseshoe orbits, in which the object librates around both points without ever approaching the planet controlling the resonance. We direct the interested reader to figures 1 and 2 of Chebotarev (1974), which illustrate tadpole and horseshoe behaviour in a simple and aesthetically pleasing manner.

Although the existence of the stable L4 and L5 points was postulated by Lagrange in the late 18[th] century, the first Trojan object, asteroid 588 Achilles, was not discovered until 1906, moving near Jupiter's L4 Lagrange point. Since then, over 3000 Jovian Trojans have been discovered, and it is thought that Jupiter's Trojan clouds may contain more objects than the main asteroid belt. The Jovian Trojans include a number of objects with high inclinations, a fact which current theories of Solar system formation are taking great pains to try to explain (Fleming & Hamilton 2000; Morbidelli et al. 2005).

Jupiter is not the only planet with asteroidal attendants. The Earth is known to have at least three "quasi-satellite" asteroids, while Mars has a retinue of at least four objects (Scholl, Marzari & Tricarico 2005). The discovery of the object 2001 QR322 (Chiang et al. 2003) meant that Neptune became the fourth planet known to be accompanied by a retinue of such objects. Since then, a further five Neptunian Trojans have been discovered, and these bodies may well represent a unique window into the formation of our planetary system, since they are believed to have moved on their current orbits since the planets settled into their current architecture. These Trojans are considered to be primordial objects, rather than temporarily captured bodies (e.g. Horner & Evans 2006), since the best-fit orbits currently available place them in an area of orbital element phase-space which has been shown to be very highly stable in previous studies of the stability of Neptunian Trojans (Nesvorny & Dones 2002). Since the discovery of these first few objects, conservative unbiased estimates of the population of the Neptunian Trojan family suggest that it houses at least as many large objects as the Jovian family (i.e. objects larger than 50-100 km in diameter), and is likely to be far more populous (containing at least an order of magnitude more objects) (Chiang et al. 2003; Sheppard & Trujillo 2006). Given that it is widely accepted that Jupiter Trojans are at least as numerous as main belt asteroids, the population of Neptune Trojans could easily outnumber that of the main belt.

To some extent, the existence of Neptunian Trojans had long been postulated (e.g. Mikkola & Innanen 1992). However, the high spread of inclinations in the small population discovered to date has caused something of a stir. Under the assumption that Neptune's Trojans, like Jupiter's, are primordial objects, stored in the Trojan clouds since the formation of the Solar system, theories which describe the formation of the Solar system must also explain the nature of these objects. Most traditional theories of planetary formation involve a fairly gentle and slow accretion of material from a cool, flat disk around the Sun (Pollack et al. 1996; Ida & Lin 2004 and references therein). Such schemes would, as a result of the dynamically cold disk[1] from which the objects form, lead one to assume that the Neptunian and Jovian Trojan populations should be confined to low inclinations (Marzari & Scholl 1998; Chiang & Lithwick 2005). Fortunately, in the light of the these observations and our ever increasing knowledge of the Solar and extra-solar planetary systems, recent years have seen the development of a number of more violent variations of cosmogenic theories. These theories, which incorporate the previously established planetary theory of planetary migration (e.g., Goldreich & Tremaine 1980; Fernandez & Ip 1984), and may even invoke the mutual scattering of the giant outer planets, attempt to explain such diverse observations as the presence of hot Jupiters (Butler et al. 1997; Masset & Papaloizou 2003), the proposed Late Heavy Bombardment (Gomes et al. 2005; Chapman, Cohen & Grinspoon 2007), the dynamically "hot" population of Jovian Trojans (Morbidelli et al. 2005), and the structure and dynamics of the Asteroid and Edgeworth-Kuiper belts (Petit, Morbidelli & Chambers 2001; Lykawka & Mukai 2008; Levison et al. 2008).

---

[1] In this work, a disk of debris is considered dynamically cold if it has never experienced significant stirring from external sources, and so mimics the very flat nature of the pre-Solar disk. Specifically, we consider disks that contain only objects with eccentricities below ~0.01 and inclinations below ~0.6° to be dynamically cold.

The unexpected discovery of four high-inclination Neptunian Trojans (objects with inclinations greater than five degrees), leads to the conclusion that there may be many more highly inclined Trojans than their low-inclination counterparts (Sheppard & Trujillo 2006), given that the discovering surveys were concentrated in the plane of the ecliptic. This proposed excess of highly inclined Neptune Trojans challenges the various mechanisms which have been proposed for the formation of Trojan objects, which invariably only produce Trojans with low orbital inclinations (e.g. Chiang & Lithwick 2005; Hahn & Malhotra 2005). This adds an important new datum to the study of Solar system formation, one which may help determine which theories are most appropriate for our Solar system (e.g., Ford & Chiang 2007; Lykawka & Mukai 2008; Levison et al. 2008). The Neptune Trojans have also been observed to possess peculiar surface colours, when compared to objects in the various classes of trans-Neptunian objects (Sheppard & Trujillo 2006). This finding adds another constraint which must be explained by future studies of Neptune's Trojans.

Given these surprising results, it is clearly important to obtain a better understanding of the formation, evolution, and dynamical behaviour of objects in the Neptunian Trojan cloud[2]. All previous work carried out to model the behaviour of the Trojans has been based on the existence of an initial population which was present by the time that Neptune had formed. That population was often based, with little stated justification, on arbitrary eccentricity and inclination distributions, or on the Jovian Trojan population (Gomes 1998; Nesvorny & Dones 2002; Kortenkamp, Malhotra & Michtchenko 2004). However, few studies followed the dynamical evolution of these objects for periods of one Gyr or more. Several authors investigated the effect of planetary migration resulting from the interaction between the giant planets and the planetesimals remaining after their formation on the stability of pre-formed Trojans (Gomes 1998; Kortenkamp, Malhotra & Michtchenko 2004). Such studies have shown that the survival rate of pre-formed Trojans is typically tens of percent and is strongly dependent on both the radial extent and the rate of Neptune's migration through the primordial planetesimal disk. Fewer studies have investigated the efficiency with which initially non-resonant objects can be captured to Trojan orbits as a result of planetary migration. Chiang et al. (2003) found that no Trojans were captured from a population of 400 objects, lying initially in a dynamically cold disk at 24.1-29.1 AU, as a result of Neptune migrating from 23.1 to 30 AU. However, Lykawka & Mukai (2008) reported that a small number of objects could be captured from the trans-Neptunian disk to Trojan orbits, and then survive for long periods, as the disk evolves over a 4 Gyr period.

In this work, we examine both the transport and capture of Trojans as a result of the smooth migration of Neptune over a significant fraction of the outer Solar system, in a manner consistent with many models of planetary formation. We compare the number of objects captured to the Trojan family by the migration of the planet to those that form in-situ, and are carried along with it, for cases of both rapid and gentle migration, and for both short (~7 AU) and long (~12 AU) migrations (i.e. for scenarios in which Neptune is initially located 23.1 or 18.1 AU from the Sun). This allows us to draw conclusions on the range and rate of Neptunian motion required to excite

---

[2] Henceforth, for brevity, the term "Trojans" will be used to refer to Neptune's Trojans.

Trojans to such high inclinations, and allow us to examine which scenario provides the best fit to the currently known Trojan objects.

The model we present represents a significant improvement on previous work, incorporating a number of new and novel features. Previous studies have typically considered a single initial value for Neptune's heliocentric distance (23 AU)[3], although work on other aspects of planet formation suggest it could easily have formed far closer to the Sun. Our work also represents an improvement of over two orders of magnitude on earlier work in the number of Trojans modelled. Furthermore, we examine how the efficiency with which Trojans are captured during migration depends on the initial orbital eccentricities and inclinations of objects in the trans-Neptunian disk, and carry out the first full study of the dynamical evolution of Trojans over the period of Neptunian migration. We consider both captured and pre-formed source populations with initially cold orbital conditions, as would be expected in the early stages of Solar system evolution. We aim to describe the Trojan cloud as a whole, with a particular focus on understanding the origin of the intriguing range of orbital inclinations represented in the known sample of objects. Previous studies were unable to address this point, reporting only individual object captures (e.g. Horner & Evans, 2006), or focussing on specific objects, so cannot be considered fully dynamical models of the Trojans (Tsiganis et al. 2005; Li, Zhou & Sun 2007; Lykawka & Mukai 2008).

In section two, we will discuss the currently known Trojans, presenting the results of simulations intended to identify their stability and general behaviour. In section three, we present the method with which runs detailing the capture and transport of Trojan objects as a by-product of Neptunian migration are constructed, before presenting a detailed analysis of the results of this work in section four. In section five, we discuss the implications of our work, and use our results to make predictions about the nature of Neptune's Trojan clouds, before drawing our conclusions, and discussing future work, in section six.

## 2 THE KNOWN NEPTUNIAN TROJANS

As of 5[th] February 2009, six Trojans have been discovered. Their orbital data are displayed below, in Table 1.

| Prov. Des. | $L_n$ | $a$ (AU) | $e$ | $i$ (°) | $\Omega$ (°) | $\omega$ (°) | $M$ (°) | $\sigma_a$ (1$\sigma$) | $\sigma_e$ (1$\sigma$) | $\sigma_i$ (1$\sigma$) | $H$ | $T_{arc}$ (days) |
|---|---|---|---|---|---|---|---|---|---|---|---|---|
| 2001 QR322 | 4 | 30.3023 | 0.031121 | 1.323 | 151.628 | 160.73 | 57.883 | 0.008813 | 0.0003059 | 0.0009417 | 7.287 | 1450 |
| 2004 UP10 | 4 | 30.2115 | 0.028449 | 1.43 | 34.799 | 358.505 | 341.278 | 0.01673 | 0.001057 | 0.002612 | 8.842 | 758.01 |
| 2005 TN53 | 4 | 30.1795 | 0.064719 | 24.986 | 9.277 | 85.716 | 287.046 | 0.01052 | 0.001583 | 0.002668 | 9.026 | 711.36 |
| 2005 TO74 | 4 | 30.1901 | 0.051836 | 5.251 | 169.375 | 302.574 | 268.101 | 0.009033 | 0.00154 | 0.002606 | 8.477 | 710.35 |
| 2006 RJ103 | 4 | 30.0772 | 0.027657 | 8.161 | 120.765 | 27.062 | 238.716 | 0.006143 | 0.0006704 | 0.0002506 | 7.438 | 796.93 |
| 2007 VL305 | 4 | 30.0423 | 0.063481 | 28.108 | 188.593 | 214.891 | 353.173 | 0.01452 | 0.0005328 | 0.002264 | 7.889 | 740.01 |

**Table 1:** List of the currently known Trojans, taken from the Asteroids Dynamic Site – AstDyS[4]. Here, $L_n$ gives the Neptunian Lagrange point about which the object librates, and $H$ the absolute magnitude of the object (the apparent magnitude it would have, observed in the V-band, if it were placed one AU from the Earth and the Sun, and displayed a full face to the Earth). $M$ gives the mean anomaly of the object (5th Feb 2009), $\omega$ gives the argument of the object's perihelion, $\Omega$ gives the longitude of its ascending node, $i$ gives the inclination of the orbit with respect to the ecliptic plane (all four angles measured in degrees of arc), $e$ the eccentricity, and $a$ the semi-major axis (AU). $\sigma_{a,e,i}$ gives the 1$\sigma$ error for the variable in question (in the appropriate units), while $T_{arc}$ gives the orbital arc covered by observations taken into account in the AstDyS orbit computation.

Two things are immediately apparent when one looks at Table 1. Firstly, all of the known Trojans librate around the leading, L4, Lagrange point. However, it seems that this result is not actually statistically significant – the surveys which have found these objects have concentrated on the leading Lagrange point, since the trailing, L5, point is currently located around the same area on the sky as the centre of our galaxy (Sheppard & Trujillo 2006). Clearly, when one is searching for faint, slow moving, star-like points, the centre of our galaxy is about the worst possible place to look! Unfortunately, since Neptune is slow moving, it may be a number of years before such observational biases are removed, and a real picture emerges of the degree of symmetry (or indeed asymmetry!) between the populations around the two Lagrange points. The second detail which is immediately obvious from Table 1 is that the objects seem to be spread in three duplets in inclination. Two objects (2001 QR322 and 2004 UP10) have low inclinations, as would be expected had they formed from a dynamically cold disk. Two Trojans lie at more intermediate inclinations (2005 TO74 and 2006 RJ103), while the final two (2005 TN53 and 2007 VL305) are highly inclined to the plane of our system. Even though only these six are currently known, it is obvious that they represent a particularly dynamic and excited population. The orbital properties of the six Trojans are illustrated in Figure 1.

---

[3] Hahn & Malhotra (2005) considered Neptune starting at 21.4 AU in their simulations. Gomes (1998) also performed a few simulations with Neptune starting at 21 AU.
[4] http://hamilton.dm.unipi.it/

After the discovery of 2001 QR322, several researchers investigated the orbital properties and stability of this Trojan. Their results indicated that it is likely that 2001 QR322 has been resident within the L4 Trojan cloud for at least 1 Gyr (Chiang et al. 2003; Marzari, Tricarico & Scholl 2003; Brasser et al. 2004). Sheppard & Trujillo (2006) went further, stating that the first four Trojans to be discovered were moving on orbits that are stable over timescales comparable to the age of the Solar system. More recently, Li, Zhou & Sun (2007) found that 2005 TN53 is also on an apparently highly stable orbit, with stability shown for a 1 Gyr period. Given that, in the current Solar system, Neptune has only an extremely small chance of capturing transient objects as Trojans (Horner & Evans 2006), and taking into account the fact that various studies have shown the currently known objects to be dynamically stable, logic dictates that they are probably primordial bodies (see Dotto et al. 2008 for more details).

In order to determine the resonant properties of the six known Trojans, we integrated the orbits of the nominal object and 100 clones of each Trojan (distributed within the $3\sigma$ orbital uncertainties for that object's semi-major axis and eccentricity, using the values given in Table 1) over 10 Myr. This allowed us to obtain the location of the centre of libration, the libration period, and the libration amplitude for each Trojan (Table 2). The libration of a given object undergoes cycles during which the resonant angle varies from minimum to maximum values, as can be seen in Figure 2. The average libration amplitude, $A$, is calculated by sequentially measuring the individual maximum and minimum displacement values for each clone over the course of the integration, then averaging over these to get the best fit to this value. Effectively, then, $A$, gives the average of the libration amplitudes experienced by all clones of the object over the 10 Myr run.

It is also interesting to calculate the exact location of the centre of libration ($C_L$) for each object considered, since the libration followed by a given Trojan is unlikely to be perfectly regular around the precise location of the L4 point. Typical values for $C_L$ will typically lie within five degrees of the nominal location of the Lagrange point (i.e. $60\pm5°$, $300\pm5°$ and $180\pm5°$ for L4, L5 and horseshoe orbits, respectively). The libration amplitudes were calculated using the *RESTICK* code (Lykawka & Mukai 2007b), with errors calculated over the dispersion of libration amplitudes obtained for the 101 bodies, and are in agreement with those obtained in previous work (Brasser et al. 2004; Lykawka & Mukai 2007a; Li, Zhou & Sun 2007). As can be seen from Table 2, all currently known Trojans are found to librate with relatively small amplitudes around the L4 point (with the exception of 2001 QR322, whose behaviour we will examine in more detail in a future paper (Horner & Lykawka 2009 (*in prep*))). When these results are compared with previous work on the stability of theoretical Trojans (e.g. Holman & Wisdom 1993; Nesvorny & Dones 2002; Marzari, Tricarico & Scholl 2003; Dvorak et al. 2007), the objects clearly fall in a region that can be considered dynamically stable, supporting the idea that they are primordial objects.

| Prov. Des. | $C_L$ (°) | $A$ (°) | $T_L$ (yr) |
|---|---|---|---|
| 2001 QR322 | 66 ± 1 | 25 ± 2 | 9200 |
| 2004 UP10 | 61 ± 1 | 12 ± 3 | 8850 |
| 2005 TN53 | 59 ± 2 | 10 ± 6 | 9450 |
| 2005 TO74 | 61 ± 2 | 13 ± 6 | 8850 |
| 2006 RJ103 | 59 ± 1 | 8 ± 3 | 8850 |
| 2007 VL305 | 59 ± 1 | 14 ± 1 | 9600 |

**Table 2:** Resonant properties of the known Trojans, obtained from calculations using *RESTICK* (Lykawka & Mukai 2007b). The values of mean libration centre ($C_L$, the distance between the mean location of the object and the position of Neptune, in degrees), mean libration amplitude ($A$, the time-averaged maximum displacement of the object from the centre of libration) and median libration period ($T_L$) are calculated from individual values obtained for the nominal object and 100 clones, after integrating their orbits for 10 Myr. The error bars show the statistical errors (at the $1\sigma$ level) resulting from averaging the libration amplitudes over the suite of 101 test particles used.

## 3 MODELLING PLANETARY MIGRATION

It is quite possible that, trapped within the details of the orbital distribution and physical properties of the Trojans, a wealth of information is preserved which can inform us of the processes which took place during the latter stages of the formation of our Solar system, in particular the way in which Neptune reached its current location and mass. The modern models that have been proposed to explain the structure of our Solar system require the giant planets to have undergone significant migration during the latter stages of their formation (Fernandez & Ip 1984; Malhotra 1995; Gomes, Morbidelli & Levison 2004; Hahn & Malhotra 2005) – in some cases in a rapid and chaotic manner (e.g. Levison et al. 2008 and references therein). Such migration is considered necessary in order to explain a number of the observed properties of our own Solar system (including the structure of the Edgeworth-Kuiper belt and the particular eccentricities and inclinations of objects locked in resonance with Neptune), together with the properties of newly discovered exoplanetary systems (e.g., Masset & Papaloizou 2003).

It is therefore clear that any study of the formation and evolution of the Trojan population must take account of the great changes in the orbital location of the planet during the final years of its formation. However, to attempt to model each possible variant for Neptunian behaviour would be hugely prohibitive, so here we present results on the behaviour of the Trojans in four representative cases. Two initial starting positions were chosen for Neptune (~18 and ~23 AU), in an attempt to bracket the minimum and maximum initial locations suggested for the planet by past work using the standard models (e.g. Lykawka & Mukai 2008 and references therein). For each of these cases, two scenarios for the migration speed were considered – one in which the planet migrated slowly, taking 50 Myr to reach its final location at ~30 AU (using an exponential-folding time of $\tau = 10$ Myr), and one with a faster rate of movement (5 Myr from start to finish, with $\tau = 1$ Myr). The integrations were carried out using the *n*-body package *EVORB* (Brunini & Melita 2002), modified to incorporate migration in such a way that Neptune's semi-major axis would vary according to

$$a_k(t) = a_k(F) - \delta a_k \exp(-t/\tau) \qquad (1)$$

where $a_k(t)$ is the semi-major axis of the planet after time $t$, $a_k(F)$ is the final (current) value of the semi-major axis, and $\tau$ is a constant determining the rate of migration of the planet. The fast and slow migration runs described above employed $\tau$ values of one and ten Myr, respectively, and the objects were followed for a period of $5\tau$ in both cases, after which the planets had reached their current locations. The index $k$ refers to the four giant planets, Jupiter ($k = J$), Saturn ($k = S$), Uranus ($k = U$) and Neptune ($k = N$). Such migration has been modelled in several previous studies (e.g. Malhotra 1995; Chiang et al. 2003; Hahn & Malhotra 2005), and represents a well accepted simplification for the migration process.

In reality, given that the migration of Neptune would involve it perturbing and displacing a vast number of smaller bodies, initially located in a broad disk beyond the orbit of the planet, varying in size from grains of dust to planetary embryos, the true migration of the planet must have been stochastic and jumpy (Hahn & Malhotra 1999; Murray-Clay & Chiang 2006). This would, in turn, be expected to lower the efficiency with which objects are captured into the many MMRs migrating ahead of the planet through $a$-space. However, given that the mass of Neptune was likely far greater than the vast majority of particles it encountered, it is fair to assume, as a first approximation, that its migration was reasonably smooth.

Our integrations took into account the gravitational influence of all four giant planets over the course of their migration. Jupiter and Saturn started each run at 5.4 and 8.6 AU respectively. For each Neptunian starting position (18.1 and 23.1 AU), additional different initial configurations were tested for Uranus, with the planet starting at locations between 12.2 and 14.7 AU (for Neptune starting at 18.1 AU), or between 14.8 and 16.6 AU (Neptune initially at 23.1 AU) (See Table 3). The migration of each planet took place over identical timescales ($5\tau = 5$ or 50 Myr, dependant on the initial $\tau$ chosen). We set the value of $\delta a_k$ so that the planets migrated from their starting locations to their current ones (in other words, Jupiter migrated inwards, while the other planets migrated outwards).

There are two potential sources for Trojan objects as Neptune marches through the outer Solar system. The first is the population of Trojans expected to form along with the planet, from material located around the two stable Lagrange points within its orbit. As the planet migrates, these objects would be carried along with it, as the 1:1 MMR sweeps outward. At the same time, it is possible that Neptune would also acquire new Trojans, as objects which grew in the trans-Neptunian region (the aforementioned disk of perturbers) are swept up by the resonance as the planet moves outwards. Thus, the reservoir of objects which formed beyond the planet provides our second source of potential Trojans. Figure 3 shows a schematic of the initial conditions used for the two distinct populations.

In order to examine the relative contribution of these two populations to the final post-migration Trojan population, clouds of massless particles were initially distributed within our simulations over a range of tadpole orbits around both Neptunian Lagrange points, together with a broad disk of objects located beyond the planet. In the two cases in which Neptune's migration was fast, $1\text{-}3\times10^4$ particles were placed in a uniform, dynamically cold, disk stretching from 1 AU beyond Neptune to a distance of 30 AU (representing the disk of primordial planetesimals), and a further 500-834 objects were spread around each of the L4 and L5 points (representing the pre-formed Trojans). In the case of slow migration, however, the capture and (in the case of N18-S1, Table 3) even the transport of pre-formed Trojans was so inefficient that the initial population had to be increased in order to obtain worthwhile statistics. The details of the four cases studied can be seen in Table 3.

The particles in the pre-formed Trojan clouds were smoothly distributed around Neptune's semi-major axis, with values varying by up to 0.1 AU on either side of the planet's initial location (for the case of N18-S1, where the ease of transport was found to be prohibitively low, the particles were distributed over a larger area, stretching from 17.8 to 18.4 AU, and 60 times more test particles were followed, in order to obtain statistically significant results). All pre-formed Trojans had initial eccentricities and inclinations in the range $0 - 0.01$, values typical of a dynamically cold disk (e.g. Hahn & Malhotra 2005 and references therein). Each object was placed with a random initial libration amplitude, $A$, such that the objects lay within ± 40º of either the L4 or L5 point. Finally, the other rotational orbital elements were randomly determined with values spanning the range $0 - 360º$.

All objects within the disk also started in a dynamically cold, unstirred state, with inclinations below 0.6º and eccentricities below 0.01 ($\sim i$). Later, we tested slightly hotter disks (one containing particles with $e < 0.05$, and a second with particles of $e < 0.1$, with inclinations determined from $e = \sin i$), in order to check how the capture efficiency changes as a function of disk excitation. All orbits were integrated over a period of $5\tau$, after which the giant planets had obtained their present day orbits after evolving according to Eq. 1. Bodies that reached heliocentric distances greater than 200 AU were removed from the calculation at that point, and not followed any further.

Finally, we used the *RESTICK* package to examine the data, detect the Trojans present at the end of the simulation, and determine their resonant properties.

| Variant code | Run | $a_{U0}$ (AU) | $a_{N0}$ (AU) | $\tau$ *(Myr)* | $N_{disk}$ | $N_{in\text{-}situ}$ |
|---|---|---|---|---|---|---|
| **N18-F** | **1** | **14.1** | **18.1** | **1** | **30,000** | **1,000** |
| | 2 | 14.1 | 18.1 | 1 | - | 1,250 |
| | 3 | 14.1 | 18.1 | 1 | - | 1,668 |
| | 4 | 12.6 | 18.1 | 1 | 10,000 | 1,668 |
| **N18-S** | **1** | **14.6** | **18.1** | **10** | **100,000** | **60,000** |
| | 2 | 14.6 | 18.1 | 10 | - | 1,668 |
| | 3 | 14.7 | 18.1 | 10 | - | 1,668 |
| | 4 | 12.2 | 18.1 | 10 | 10,000 | 1,668 |
| **N23-F** | **1** | **16.1** | **23.1** | **1** | **30,000** | **1,000** |
| | 2 | 16.1 | 23.1 | 1 | - | 1,668 |
| | 3 | 14.8 | 23.1 | 1 | 10,000 | 1,250 |
| | 4 | 14.8 | 23.1 | 1 | - | 1,668 |
| **N23-S** | **1** | **16.2** | **23.1** | **10** | **80,000** | **1,000** |
| | 2 | 16.2 | 23.1 | 10 | - | 1,668 |
| | 3 | 14.8 | 23.1 | 10 | - | 1,668 |
| | 4 | 15.1 | 23.1 | 10 | - | 1,668 |
| | 5 | 14.8 | 23.1 | 10 | 10,000 | 1,250 |

**Table 3:** Model parameters. F = Fast migration, S = Slow migration. $a_{U0}$ and $a_{N0}$ = initial semi-major axis for Uranus and Neptune, respectively. $N_{disk}$ and $N_{in\text{-}situ}$ refers to the initial number objects used in the disk beyond Neptune and as pre-formed Trojans. The four principle runs are highlighted in **bold** text. Single dashes (-) indicate runs in which only pre-formed Trojans were considered, and so no disk objects were present.

## 4 RESULTS

By the end of planetary migration, a significant number of objects remained as Trojans, with origins both in the pre-formed clouds and the trans-Neptunian disk. The full gamut of Trojan behaviour was displayed, with tadpole Trojans around both L4 and L5, and horseshoe objects. Through use of *RESTICK*, the distribution of these objects was obtained in both element space ($a$-$e$-$i$) and resonant properties. In particular, we obtained information for each object on every individual resonance capture event, and results for the integration as a whole. For the individual resonance captures, we calculated the duration of the capture, the type of libration (L4, L5 or horseshoe) and the orbital elements of the object. For the integration as a whole, we obtained the total number of resonant captures, the total time spent in resonance, the number of transitions between different libration styles, the number of periods spent as a horseshoe, L4 or L5 object, and the total time spent in each of these categories.

In brief, the Trojans that had been captured from the initially cold disk displayed a wide range of elements, with eccentricities ranging from 0 up to 0.35, and inclinations from 0 to 50º, while those which had been formed with Neptune and were then transported along with it typically had small eccentricities and inclinations ($e < 0.1$, $i < 5$º). The one exception to this behaviour was the highly unstable case of slow migration from 18 AU (N18-S1-3), which resulted in the pre-formed Trojans being excited to orbits ranging up to eccentricities of 0.35 and inclinations of 40º. In terms of resonant properties, both pre-formed and captured Trojans yielded libration amplitudes from virtually zero (though captured objects rarely attained amplitudes less than ten degrees) to ~60-70º for L4/L5 orbits (in the limit of leaving the tadpole orbit) and 150-170º for horseshoe orbits.

| Variant | Run | $C_d$ (%) | $N_{Hd}$ | $N_{L4d}$ | $N_{L5d}$ | $R_p$ (%) | $N_{Hp}$ | $N_{L4p}$ | $N_{L5p}$ | $f_{RC}$ |
|---------|-----|-----------|----------|-----------|-----------|-----------|----------|-----------|-----------|----------|
| N18-F | 1 | 0.750 | 174 | 26 | 25 | 54.6 | 208 | 141 | 197 | ~73 |
| N18-S | 1 | 0.120 | 72 | 28 | 20 | 0.148 | 37 | 23 | 29 | ~1.2 |
| N23-F | 1 | 1.09 | 268 | 27 | 31 | 96.5 | 242 | 352 | 371 | ~89 |
| N23-S | 1 | 0.129 | 50 | 27 | 26 | 50.3 | 12 | 242 | 249 | ~390 |

**Table 4:** The statistical results obtained from our principle runs. For each setup, we detail the capture efficiency of Trojans from the trans-Neptunian disk (stretching to 30 AU, $C_d$). In addition, $R_p$ gives the retention fraction of pre-formed Trojans (the fraction of those objects which started the simulations around the L4 and L5 points which are also Trojans at the end of the simulation). $N_H$, $N_{L4}$ and $N_{L5}$ give the number of particles from remaining on horseshoe and tadpole orbits at the end of the simulations (with subscript $_d$ denoting those captured from the main disk, and $_p$ denoting those which were pre-formed). Finally, $f_{RC}$ gives the ratio of the probability of pre-formed Trojans being retained ($R_p$) to the likelihood of objects from the disk being captured ($C_d$).

| Variant | Run | $C_d$ (%) | $R_p$ (%) | $N$ |
|---------|-----|-----------|-----------|-----|
| N18-F | 2 | - | 38 | 475 |
| | 3 | - | 47 | 784 |
| | 4 | ~0.4 (37) | 98 | 1635 |
| N18-S | 2 | - | 0 | 0 |
| | 3 | - | <0.1 | 1 |
| | 4 | ~0.2 (24) | 98 | 1635 |
| N23-F | 2 | - | 87 | 1451 |
| | 3 | ~0.3 (28) | 70 | 875 |
| | 4 | - | 83 | 1384 |
| N23-S | 2 | - | 33 | 550 |
| | 3 | - | 72 | 1201 |
| | 4 | - | 36 | 600 |
| | 5 | ~0 (2) | 71 | 888 |

**Table 5:** The results of additional, less detailed data output runs carried out to examine the effect changes to the initial conditions (degree of initial Trojan excitation, initial positions of Uranus and Neptune) had on the results for each variant studied. As with Table 4, $C_d$ details the capture efficiency of particles from the disk of objects spread from just beyond the initial location of Neptune out to 30 AU, $R_p$ gives the fraction of pre-formed Trojans retained at the end of the simulation, and $N$ gives the number of Trojans present at the end of the run. Single dashes (-) indicate runs in which only pre-formed Trojans were considered, and so no disk objects were present.

As can be seen in Tables 4 and 5, the stability of pre-formed Trojans during planet migration was greatly dependent on both the initial heliocentric distance and the rate of migration, with survival rates ranging from approximately 98% (18-F4; and 18-S1) to total loss (18-S2 and 18-S3). However, the survival rates for the cases where Neptune started at 23.1 AU (several tens of percent) are in agreement with past work (Gomes 1998; Kortenkamp, Malhotra & Michtchenko 2004).

Finally, a relatively large number of Trojans surviving on horseshoe orbits were obtained in all runs (See Table 4 for details). However, because objects on such orbits are typically unstable, one can expect these objects not to survive over the age of the Solar system, and therefore such objects are not examined in any great detail in this work.

## 4.1 BEHAVIOUR OF PRE-FORMED TROJANS

An examination of the behaviour of the pre-formed Trojans over the period of Neptune's migration allows us to determine the effect that the various model parameters considered have on the final distributions of these objects. Tables 4 (which shows the main runs carried out) and 5 (which details subsidiary runs carried out with lower output detail) present the number of objects which survive the duration of Neptune's migration as Trojan objects, expressed as a percentage of the initial population. For the four main runs, Table 4 also presents the final distribution of the surviving objects among the three categories of Trojan object – L4, L5 and Horseshoe objects.

It is initially clear, from Table 4, that the retention of pre-formed Trojans is greatly dependant on the rate of Neptune's migration. For both initial values of Neptune's semi-major axis (18.1 and 23.1 AU), we found that far more objects were retained for rapid migration than for slow migration. This initially surprising result may be explained by the fact that a faster migration rate both allows less time for objects to escape from the 1:1 MMR, and also means that objects carried within that resonance will spend less time in de-stabilising secondary resonances[5] as they sweep through the outer Solar system (e.g., See Kortenkamp, Malhotra & Michtchenko 2004).

Table 4 also reveals that, for a given migration speed, the shorter the range of Neptune's migration, the greater the retention of pre-formed Trojans. This result is perhaps less unexpected than the first, since fewer harmful secondary resonances will be encountered before the objects settle in their final locations. Therefore, it is less likely that there will be widespread disruption of the pre-formed Trojan clouds.

More specifically, from Table 4, the scenario which offered the best retention of pre-formed Trojans was that in which Neptune migrated rapidly from 23.1 AU. In this case, 96.5% of objects were retained for the course of the planet's migration. At the other extreme, the scenario with the worst retention rate was that in which Neptune migrated slowly from 18.1 AU to its final location. Here, so few objects were retained that the initial population had to be greatly enhanced (to 60000 objects, see Table 3) in order to obtain reasonable statistics on the nature of surviving objects, with a retention rate of just 0.148%. It is also interesting to note that, in the case of slow migration from 23.1 AU, very few objects successfully make the transition between tadpole and horseshoe type orbits, while for the other three scenarios the final number of horseshoe objects is comparable to those on tadpole type orbits.

Figures 4 and 5 show the distribution of the post-migration Trojan objects for each of the four cases detailed in Table 4, showing the eccentricity and inclination of their orbits (figure 4) and their libration angles and inclinations (figure 5). A few details are immediately clear from these plots.

In three out of the four scenarios considered, the initially dynamically cold swarms of pre-formed Trojans are barely excited in inclination or eccentricity. Even pre-formed Trojans forced onto horseshoe orbits retain their initially low eccentricities and inclinations. In other words, in these scenarios, the pre-formed Trojan clouds are insufficient to explain the observed properties of the modern-day Trojan population. In the case where Neptune migrates slowly outward from 18.1 AU, the pre-formed Trojans are heavily disrupted, being excited to both high inclinations and eccentricities by severe gravitational perturbations from Uranus and Neptune, leading to an almost complete loss of objects from the pre-formed clouds. However, many of these objects continue to be forced outwards by the planetary migration, and are later recaptured as Trojans, leading to the great similarities which can be seen in the distributions of "pre-formed" and captured populations in that case. Figure 6 shows a typical example of such behaviour. Note how the particle is initially ejected from the pre-formed cloud, then follows the outward migration of Neptune, hopping between a number of short-lived interior and exterior MMRs, before finally being re-captured as a Trojan. A more detailed explanation of the behaviour of this particular object is given in the caption for Figure 6.

---

[5] Secondary resonances often involve commensurabilities between the characteristic libration frequency of objects within a given MMR and other frequencies, such as the libration and circulation frequencies of Trojans near that resonance (Murray & Dermott 1999; Kortenkamp, Malhotra & Michtchenko 2004).

Slow migration typically leads to fewer tadpole survivors with large libration amplitudes (> 60°) than fast migration. This may, however, be a result of the longer timescales considered for the slow migration runs – objects with the greatest tadpole libration amplitudes have historically been shown to be the least stable (e.g., Nesvorny & Dones 2002), so this feature may simply be a result of the population of such objects decaying before the end of the run for the slow migration cases. A detailed analysis of the post-migration stability of Trojans obtained in this paper will be given in a future work (Lykawka et al. 2009b (*in prep*)).

In Table 5, we present the results of secondary runs carried out in order to examine the various subtle features exposed by the key runs (these extra runs are runs 2-4/5 for each of the cases, as detailed in Table 3). In order to obtain these results on a reasonable timescale, a less detailed output was chosen, which precluded the depth of analyses given to the key runs. The results do, however, highlight a couple of important features. Firstly, a number of these subsidiary runs invoked different initial planetary architectures, with the planet Uranus starting in a different location (18F-4; 18S-3,4; 23F-3,4; 23S-2-4). When the retention fraction of pre-formed Trojans from these runs is compared to that for the main runs, it is clear that the initial architecture of the system used can play an equally important role to the nature and range of the migration of Neptune. The best two examples of this can be seen in runs 18F-4 and 18S-4, where placing Uranus further from Neptune at the start of the simulation has a drastic effect on the number of retained objects (in both cases increasing the retention fraction to over 90%). The initial conditions for runs 18S-1, 18S-2 and 18S-3 have Uranus and Neptune beginning their motion very close to their mutual 3:4 MMR, which may have induced strong instabilities through overlapping of this resonance with the Trojan's 1:1 resonant motion (e.g. Kortenkamp, Malhotra & Michtchenko 2004). This shows the great importance that mutual resonance crossing events can have in destabilising the pre-formed Trojans. This finding also confirms the fundamental role that the location of Uranus played during the migration of the outer planets, being a major influence on the degree of disruption suffered by the Trojan population (Gomes 1998). A detailed analysis of this phenomenon is beyond the scope of this work, but it will be the subject of a forthcoming paper (Lykawka et al. 2009a (*in prep*)).

It should, at this point, be noted that the loss of objects from the pre-formed Trojan cloud was not uniform across those clouds. Indeed, there was a clear link between the initial libration amplitude, $A$, of the objects and their survival efficiency. Objects with larger initial libration amplitudes were significantly less stable, over the course of planetary migration, than those which experienced smaller scale libration. As discussed above, this is far from unexpected – the greater the scale of libration, the less stable an object would be expected to be – and so it is natural that the clouds would effectively be whittled away from the outside inwards. However, this effect was only noticeable for libration amplitudes beyond 30-35° - the transportation efficiency of objects with smaller libration amplitudes than this showed little dependence on the initial $A$.

## 4.2 BEHAVIOUR OF CAPTURED TROJANS

Tables 4 and 5 also detail the efficiency with which the migrating Neptune captured objects from the trans-Neptunian disk. In each of the four main cases (detailed in Table 4), the capture rate observed was actually surprisingly high (between 0.12% and 1.1%), when one considers the amount of scattering that objects in this region must experience as they are stirred by the migrating planet. As for the pre-formed Trojans, the efficiency with which the trans-Neptunian objects are captured is strongly dependent on the speed of the planet's migration, with cases of fast migration being almost an order of magnitude more efficient in Trojan capture than their slower counterparts. One surprising feature of these capture runs, when compared to the behaviour of pre-formed Trojans (discussed earlier) is that the capture efficiency observed in the run detailing slow migration from 18 AU was almost identical to that for the case of slow migration from 23 AU. Clearly, the highly destabilising event which affected the pre-formed cloud of Trojans had little effect on the efficiency with which objects were captured from the trans-Neptunian cloud. This adds further weight to the argument that the destabilisation occurred early in the migration of Neptune, and was only effective for a short period – after which the efficiency with which objects were captured from the trans-Neptunian disk was unaffected.

Also obvious from Table 4 is the fact that the captured Trojans move on predominantly horseshoe type orbits (an eight- to ten-fold excess in the case of fast migration, compared to a two- to three-fold excess for slow migration). Nevertheless, it is interesting to observe that a significant number of objects were captured onto the theoretically more stable tadpole orbits as the result of planetary migration.

The distribution of captured Trojans at the end of the key runs is shown in the right-hand panels of Figures 4 and 5. In the three cases for which the pre-formed Trojans remain un-excited, there is a clear difference between their distribution and that of the captured Trojans. Although a small fraction of the captured Trojans attain low eccentricities and inclinations, the Trojan populations remain essentially distinct – captured objects move on dynamically hot orbits, while pre-formed objects remain dynamically cool. The only exception to this rule is the unusual case of 18AU-S1, in which it seems likely that mutual resonant events involving Uranus and Neptune have a hugely de-stabilising effect on the pre-formed Trojan population. Even in this case, however, it is apparent that the captured Trojans result in fewer objects on low excitation orbits ($e < 0.05$, $i < 5º$), while there remains an undisrupted relic of the pre-formed Trojan population in that region. From Figure 5, it is clear that captured Trojans show no real correlation between libration amplitude and orbital inclination, with the interesting result that a number of objects can be captured onto orbits very close to the L4 and L5 point. Presumably, orbits with lower libration amplitudes would be more stable than those with high libration amplitudes, and so such objects could potentially be the source of the known high-inclination Trojans (objects with $i > 5º$, which all currently have libration amplitudes less than 15 degrees – see Tables 1 and 2).

In order to check whether the capture efficiency of trans-Neptunian disk objects to the Trojan clouds was dependent on the initial dynamical state of the disk, two additional small scale runs were carried out for each of the scenarios in which migration began at 18.1 AU. The first such run assumed a slightly "hotter" disk than that used in the key run, with eccentricities up to 0.05 (and using $e = \sin i$), and the second assumed a disk that was hotter still, with eccentricities up to 0.1. For both these scenarios, it was found that the capture efficiency was essentially unaltered. The two additional runs for N18-F yielded $C_d$ values of 0.5% ($e < 0.05$) and 0.6% ($e < 0.1$) (similar to the $C_d$ ~ 0.7% achieved in the main N18-F run), and those for N18-S gave $C_d$ values of <0.1% in both cases (again, which is the same as the ~0.1% obtained in the main N18-S simulation).

| Variant | Run | $C_c$ (%) | $C_{UN}$ (%) |
|---------|-----|-----------|--------------|
| N18-F | 1 | 0 | 0.6 (30) |
| N18-S | 1 | <<0.1 | 0 |
| N23-F | 1 | 0 | 0.4 (21) |
| N23-S | 1 | <<0.1 | 0 |

**Table 6:** The results of additional simulations carried out to examine the capture efficiency of Trojans from the extended disk (5000 particles distributed in the range 30-45 AU, $C_c$), and a disk of 5000 objects initially placed between the orbits of Uranus and Neptune ($C_{UN}$) for our principle runs.

Table 5 also details subsidiary runs which examined the effects of different Uranus-Neptune architectures on the efficiency of capture from the trans-Neptunian disk (runs 18F-4, 18S-4, 23F-3 and 23S-2). The small number of captures observed in these runs prevent us from drawing too many conclusions on the effect of planetary architecture on the capture of trans-Neptunian objects as Trojans during migration, although the initial capture efficiencies seem comparable to those obtained in the key runs.

Finally, Table 6 details the results of additional runs which were carried out to examine the efficiency with which Neptune could capture Trojans from an extended disk (stretching from 30 – 45 AU), and the efficiency with which objects were captured from a swarm of objects that were initially distributed between the orbits of Uranus and Neptune (a *cis*-Neptunian disk[6]). Each disk was populated with 5000 test particles on dynamically cold ($e \sim i < 0.01$) orbits, and was followed until the migration of the planets had stopped. We found that the contribution of bodies from the extended disk to the Trojan cloud was negligible, and so it seems highly unlikely that this region contributed more than a tiny fraction of the current Trojan population, unless some event caused it to become significant excited before the end of planetary migration. In cases of slow migration, we observed no captures from the *cis*-Neptunian disk. However, for cases where the migration was fast, the capture efficiency from this region was found to be approximately 0.5%, comparable to the efficiency with which objects were captured from the trans-Neptunian region. Although the trans-Neptunian disk would span a much greater area of the Solar system, and therefore contain a significantly greater population of objects, it is clear that, at least for rapid planetary migration, the capture of objects which begin on orbits interior to that of Neptune could provide a significant contribution to the final population of captured Trojans.

## 5 DISCUSSION

In previous work (Sheppard & Trujillo 2006), the existence of the high-$i$ component of the Trojan population was taken as evidence that capture mechanisms played a vital role in the creation of the Trojan population. However, our results have shown that, in certain situations, it is feasible that mutual interactions between Uranus and Neptune over the course of their migration can perturb pre-formed Trojans leading to their temporary ejection from the Trojan cloud (Fig. 6). Such objects can then acquire highly inclined and eccentric orbits as a result of successive close encounters with Uranus and Neptune, before being recaptured as Trojans prior to the end of planet migration. This does, however, require the two planets to undergo mutual resonant events (such as the crossing of their mutual 3:4 MMR, or the action of secondary resonances (Kortenkamp, Malhotra & Michtchenko 2004)). In such cases, the great bulk of pre-formed Trojans would be ejected, with the relic population being almost indistinguishable from that which could be captured as a direct result of Neptune's migration through a disk of planetesimals. This result reinforces the belief that the mutual locations of Uranus and Neptune can play a pivotal role on the stability of Trojans formed during and after the assembly of both planets.

When Uranus and Neptune are initially situated in such a way that mutual resonant events do not happen, then it is impossible for pre-formed Trojan objects alone to explain the observed distribution of Neptune Trojans. The migration of Neptune, without such events, does not impart any measurable increase in the eccentricities or inclinations of the pre-formed Trojan population, and so a second source of material is required in order to explain the highly inclined Trojans. In cases such as these (which make up the great majority of theorised starting points for planetary migration (Hahn & Malhotra 2005; Lykawka & Mukai 2008), we concur that the existence of a large highly-inclined population of Trojans requires a significant capture of material from the trans-Neptunian debris disk. The capture efficiency of material during Neptune's migration is found to be surprisingly high, considering the highly disruptive effect the planet's motion has on the disk beyond the planet – capture rates of order 1% (for fast migration) or 0.1% (for slow migration) initially look low until one remembers that the total mass of material located between the location at which proto-Neptune was formed and its current location could easily have been measured in tens of Earth masses (Kenyon et al. 2008). Such capture efficiencies, then, allow the acquisition of a huge number of small bodies on Trojan orbits, such that, even if the great majority of such orbits are dynamically unstable on Gyr timescales, enough objects should remain at the current day to represent a significant contribution to the observed population. We are currently in the process of simulating the long term stability of the Trojan clouds formed as a result of Neptune's migration, and will present a detailed analysis of the results in a future work (Lykawka et al. 2009b (*in prep*)).

---

[6] This disk was created so that, at the start of the simulation, its inner edge was 1 AU beyond the orbit of Uranus, and the outer edge was 1 AU within the orbit of Neptune.

It is clear, therefore, that the existence of a highly inclined component to the Trojan population cannot be used to constrain their source population. Furthermore, since our results suggest that Trojans could arise solely by capture from the disk during a smooth Neptune migration phase, the known high-*i* Trojans cannot be used to directly infer that the planets in the outer Solar system must have experienced turbulent and disruptive resonant events (either resonant interactions and mutual perturbations between Uranus and Neptune (e.g., Gomes 1998; Kortenkamp, Malhotra & Michtchenko 2004) or mutual gravitational scattering between these planets (e.g., Levison et al. 2008)).

Given the observed capture efficiencies for objects located on orbits initially beyond the orbit of Neptune, it seems that the most reasonable explanation for the highly excited population of Trojans is simply that it is a direct result of capture during the planet's migration. Though each of our models is capable of reproducing some aspects of the known Trojans, determining which provides the best fit to the true formation of the Solar system is far more difficult, given the paucity of available observational data. It is worth noting, however, that those simulations which invoked rapid migration of Neptune (runs 18AU-F and 23AU-F) produced few objects with inclinations greater than 20º, the majority of which were located on orbits with relatively large libration, which implies that they would be particularly unstable on Gyr timescales (Fig. 4 and 5). Given that the six known Trojans feature two objects with high inclinations and comparatively small libration amplitudes (Tables 1 and 2) that appear to be dynamically stable, we suggest that the rapid-migration models face significant problems when attempting to explain the whole observed Trojan population. On the other hand, the results of models featuring more gradual planet migration result in significantly better inclination and libration amplitude distributions, when compared with the observed objects. This suggests that planet migration within the outer Solar system operated at a sedate, rather than hectic, pace. This conclusion is in agreement with the migration timescales found by Gomes et al. (2004) and Nesvorny et al. (2007). Of the two slow migration models considered (18AU-S and 23AU-S), it seems that a more extended migration (18AU-S) is significantly more successful in producing objects with a suitable combination of highly inclined orbits with small libration amplitudes and moderate eccentricities (<0.15), which are like those observed in the known Trojan population. However, it is vital that long term investigations of the resulting Trojan clouds are carried out in order to support the conclusion that a slow, extended migration of Neptune best explains the observed Trojans, and we are currently in the process of carrying out such work, which will be reported at a later date.

Our examination of the fate of Trojans formed in-situ has show that, depending on the initial conditions used, they could contribute to the observed Trojan population in one of two ways. In one case, the unusual scenario 18AU-S, pre-formed Trojans are excited (often through a process of ejection and re-acquisition) to orbits resembling those of Trojans captured from the trans-Neptunian disk (i.e. $e < 0.35$, $i < 60º$). In the other cases, however, where there is a paucity of significant planet-induced instabilities, the pre-formed Trojans result in a dynamically cold ($e < 0.1$, $i < 5º$) population by the completion of migration. This, therefore, could provide a key observation allowing theorists to determine whether the true migration of Neptune had involved significant interactions with the other planets, or had been more placid in nature. However, it is important to note that the relative importance of the pre-formed Trojans, when compared to those captured during Neptune's migration, will be heavily influenced by the size of the initial population of pre-formed Trojans assembled prior to onset of Neptune's outward migration. This is still very poorly constrained (Chiang & Lithwick 2005). In particular, although Chiang & Lithwick (2005) showed that accretion was feasible around the leading and trailing Neptunian Lagrange points, the efficiency with which objects can be captured to the Trojan clouds from external reservoirs would appear to suggest that such pre-formed Trojans might be greatly outnumbered by their captured brethren[7]. The contribution of pre-formed Trojans may also have been negligible if the dynamical conditions around the L4 and L5 points were hostile to accretion (for example, with objects there excited so that destructive collisions dominated over accretive encounters), preventing the formation of any significant number of objects. Finally, if the migration of the planets led to significantly more disruptive dynamical conditions for the Trojans than those considered in this work (such as could be imagined in scenarios which invoke close encounters between the giant planets themselves), this would clearly cause the loss of most, if not all, of these objects, again rendering their contribution to the final Trojan population negligible.

Our results suggest, therefore, that models of the formation of Neptune's Trojans can be broken down into two broad scenarios.

1) Non-chaotic dynamical evolution.
Here, as the planets migrate to their final locations, Neptune undergoes no significant resonant interactions with the other outer planets. In such cases, the retention of pre-formed Trojans is significantly more efficient (by two orders of magnitude) than the capture of objects from other reservoirs (Tables 4, 5 and 6). In such scenarios, therefore, in order to explain the observed Trojan distribution (and, in particular, to avoid creating an excess of low-$i$ objects when compared with the observed population), we can conclude that the initial population of pre-formed Trojans must have been significantly smaller than that of objects located in the trans-Neptunian disk of planetesimals. Indeed, from a simple comparison of the volume and surface density of material initially in these regions prior to migration, we would suggest that the amount of mass initially located on pre-formed Trojan orbits would have been at least three orders of magnitude less than that in the disk beyond Neptune (a conclusion in agreement with the accretion model of Chiang & Lithwick 2005).

2) Chaotic dynamical evolution.
In cases where the migration of the giant planets is significantly more chaotic (such as in our key 18AU-S case), the fraction of retained pre-formed Trojans is comparable to the probability that objects will be captured from the trans-Neptunian disk (Table 4). In addition, the resulting distribution of Trojans in $a$-$e$-$i$ space is such that, with the exception of a small residue of dynamically cold objects with an origin in the pre-formed clouds, there is little appreciable difference between the different potential formation mechanisms. In such scenarios, we can place no constraints on the relative sizes of the initial populations in the different formation areas.

As we have seen in our results, the effect of such chaotic periods on the transport of pre-formed Trojans can lead to great variation in both the survival of such objects, and the final distribution of those that remain in the Trojan cloud. If a significant concentration of Trojans is found on dynamically cold orbits, then this would clearly suggest that the migration of Neptune had been unhindered by chaotic interactions between the outer planets, since the only way to produce such a distribution of objects is for that planet to undergo gentle and non-chaotic migration. We found no situation where the capture of objects to the Trojan cloud could produce an excess of such low excitation objects – indeed, if anything, captured objects tend to avoid the region of lowest excitation ($e < 0.02$ and $i < 4°$). It is the existence of the dynamically cold Trojans, rather than the excited population, that provides the strongest constraint on the formation of the outer Solar system.

To summarise – the observed Trojan population cannot be explained by the presence of pre-formed Trojans alone. Depending on the initial conditions used, it is possible to produce populations of

---

[7] Given the uncertainties in the local accretion modelling, the number of surviving Trojans formed in-situ may be such that they resemble the numbers of Trojans captured from the disk in the same size range. This is plausible even if these objects possess various inclinations, because their long-term stability depend little on the latter (e.g., Dvorak et al. 2007).

Trojans which mimic that observed today (with some failings which may be explained when the long term evolution of the Trojans is considered after migration) through either a combination of pre-formed and captured objects, or even by captured objects alone. Unfortunately, the population of known Trojans is currently too small for their orbital distribution to be used to distinguish further between the different scenarios for their formation.

Future observations by missions such as *Herschel (*Mueller et al. 2009*)* will carry out the first detailed observational studies of the Neptune Trojan population, allowing better determination of the sizes and physical and chemical compositions. In addition, future very-wide-field survey projects (such as Pan-STARRS (Jewitt 2003) and the LSST (Ivezic et al. 2008)) will lead to at least an order of magnitude improvement in the size of the catalogue of known Trojans, revealing the current orbital structure of the family in more detail. Such work will provide important additional constraints that will help to determine which of the formation scenarios best fits the modern Trojans. Should the Trojans originate from two highly disparate populations, it is likely that this will be reflected in these observations, with the dynamically cold and dynamically hot populations displaying significant differences. Equally, if all (or at least, the great majority) of the Trojans have origins in the same population, one would expect observational programmes to reveal the objects to be far more homogeneous in nature. It should be noted that observations of four of the known Trojans suggest they have broadly similar colours, which would suggest that the latter of these scenarios is more likely, although we believe more observations and detailed analysis are necessary in order to draw any firm conclusions.

## 6 CONCLUSIONS AND FUTURE WORK

We have carried out the first detailed dynamical simulations of the transport and capture of objects to Neptune's Trojan population as a function of Neptunian migration. Our results show that both the rate and range of migration can play an important role in determining the survival fraction for pre-formed Trojans, and the capture probability for objects originating elsewhere. In addition, it is clear that the nature of the migration (smooth and free from strong perturbations by the other planets vs. chaotic and turbulent) can play a significant role in shaping the final distribution of Neptune Trojans. In particular, we found that the transport of pre-formed Trojan populations could be highly disrupted by mutual resonant perturbations between Uranus and Neptune, leading to the almost complete loss of a pre-formed population. However, a small, but significant, fraction of this lost population was eventually re-captured to the 1:1 MMR on orbits stable until the end of migration. Such re-captured orbits exhibited a much wider range of orbital elements than their brethren that were transported unperturbed, with typical eccentricities as high as 0.35, and inclinations up to 40º. Indeed, such recaptured Trojans were effectively indistinguishable from those captured directly from the trans-Neptunian disk. Conversely, in scenarios which did not result in such perturbations occurring for any prolonged period, we found that pre-formed Trojans were retained with an efficiency between ~30-98% (two orders of magnitude higher than that with which this objects were captured to the Trojan region from other reservoirs), and remained with dynamically cold orbital conditions, $e < 0.1$, $i < 5º$.

The efficiency with which objects were captured to the Trojan population, during migration, from the trans-Neptunian disk (which ranged between approximately 0.1% and 1% over the different scenarios) was significantly lower than the retention rate of pre-formed Trojans in the non-disruptive cases. However, this capture rate is sufficiently high that, given that many Earth-masses of material likely accreted to form planetesimals beyond Neptune's initial location, the Trojan population could well be dominated by captured objects, even in those cases with the most efficient transport of pre-formed objects. In every scenario considered, the orbits of captured Trojans covered a broad range of eccentricities and inclinations. Typically, values ranged up to eccentricities of 0.35, and inclinations of 40º (though we should note that capture during faster migration runs typically led to captured populations with slightly cooler characteristics, with few objects exceeding an inclination of ~20º). It is therefore clear that, taking our results as a whole, the highly inclined component of the Neptune Trojan population can be explained reasonably well within the framework of our models. However, of four models considered, we found that the one which best mimicked the observed distribution of Neptune Trojans involved the relatively slow migration of Neptune over an extended distance (from an initial location of 18.1 AU). Faster migration, or migration from just 23.1 AU, resulted in distributions of objects which failed to reproduce key features of the observed population.

In future work, we will follow the evolution of the Trojan clouds resulting from the various scenarios considered in this work in great detail, examining the behaviour of tens of thousands of objects on a Gyr timescale in order to better ascertain which scenarios would result in the best fit to the Trojans observed today. We will also examine the effect of a wider range of planetary architectures on pre-formed Trojan clouds.

We believe that future observational campaigns (from missions that will obtain physical and chemical information on the nature of the Trojans, such as *Herschel* (Mueller et al. 2009), to those which will greatly increase the known sample of Trojans, such as Pan-STARRS (Jewitt 2003) and the LSST (Ivezic et al. 2008), will play a key role in determining which scenarios for planetary migration are a good fit with the observed Trojan population, and encourage observers actively searching for more Trojan objects to avoid limiting their observations to the region close to the ecliptic – the nature and relative population of objects on high-inclination may prove a critical test not only for theories of Trojan formation, but also for models of Neptune's formation, planetary migration and the origin and evolution of the outer Solar system.

**ACKNOWLEDGEMENTS**
We would like to thank an anonymous referee for a number of helpful and detailed comments, which allowed us to improve the overall presentation and flow of this work. PSL and JAH gratefully acknowledge financial support awarded by the Daiwa Anglo-Japanese Foundation and the Sasakawa Foundation, which proved vital in arranging an extended research visit by JAH to Kobe University. PSL appreciates the support of the COE program and the JSPS Fellowship, while JAH appreciates the ongoing support of STFC.

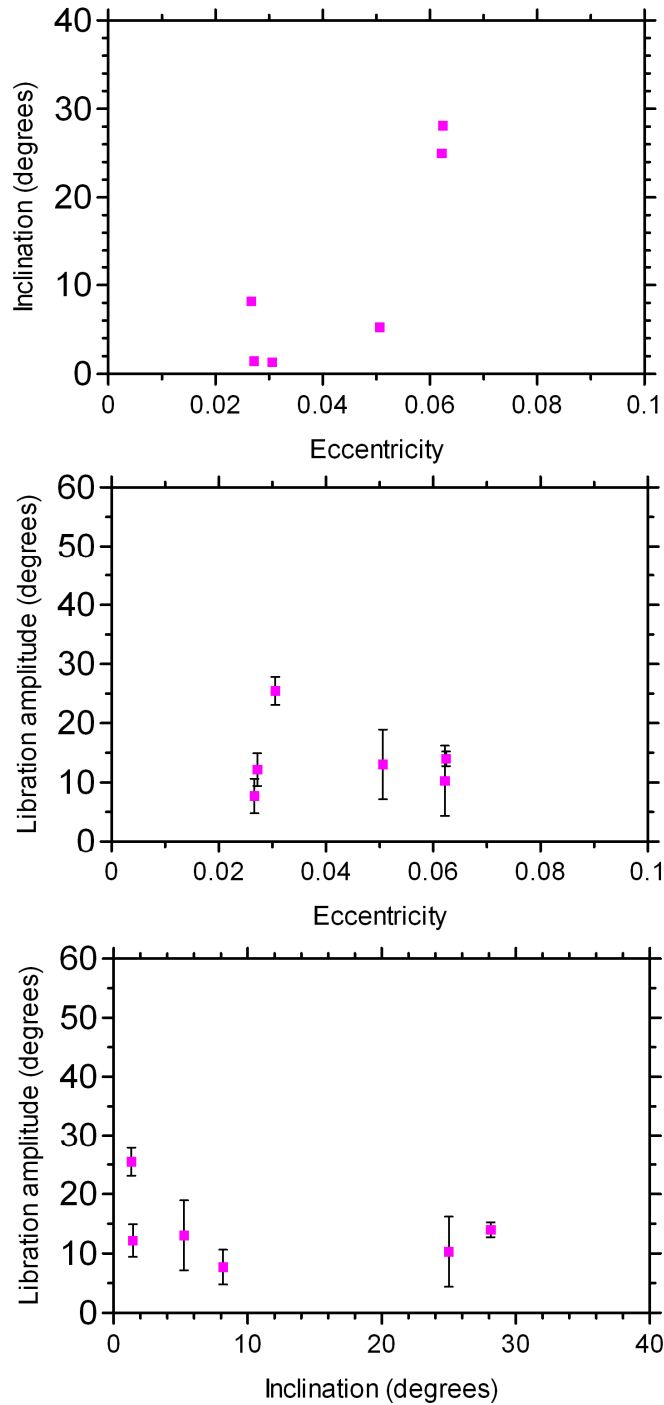

**Figure 1:** General properties of the six currently known Neptunian Trojans. Top: eccentricity vs. inclination (°). Middle: eccentricity vs. libration amplitude (°). Bottom: inclination (°) vs. libration amplitude (°). The observational data was taken from the AstDyS database on the 24th April 2008. All six Trojans orbit in the vicinity of Neptune's L4 point. The libration amplitudes were averaged over individual values calculated for the nominal object plus 100 clones over integrations following their orbits 10 Myr into the future. Here, the libration amplitude refers to the maximum angular displacement from the centre of libration during the object's resonant motion. Libration amplitudes were calculated using the *RESTICK* code (Lykawka & Mukai 2007b). The error bars show the statistical errors (at the $1\sigma$ level) resulting from averaging the libration amplitudes over the suite of 101 clones used. Details of the resonant properties of the known Trojans, obtained from these integrations, are shown in Table 2.

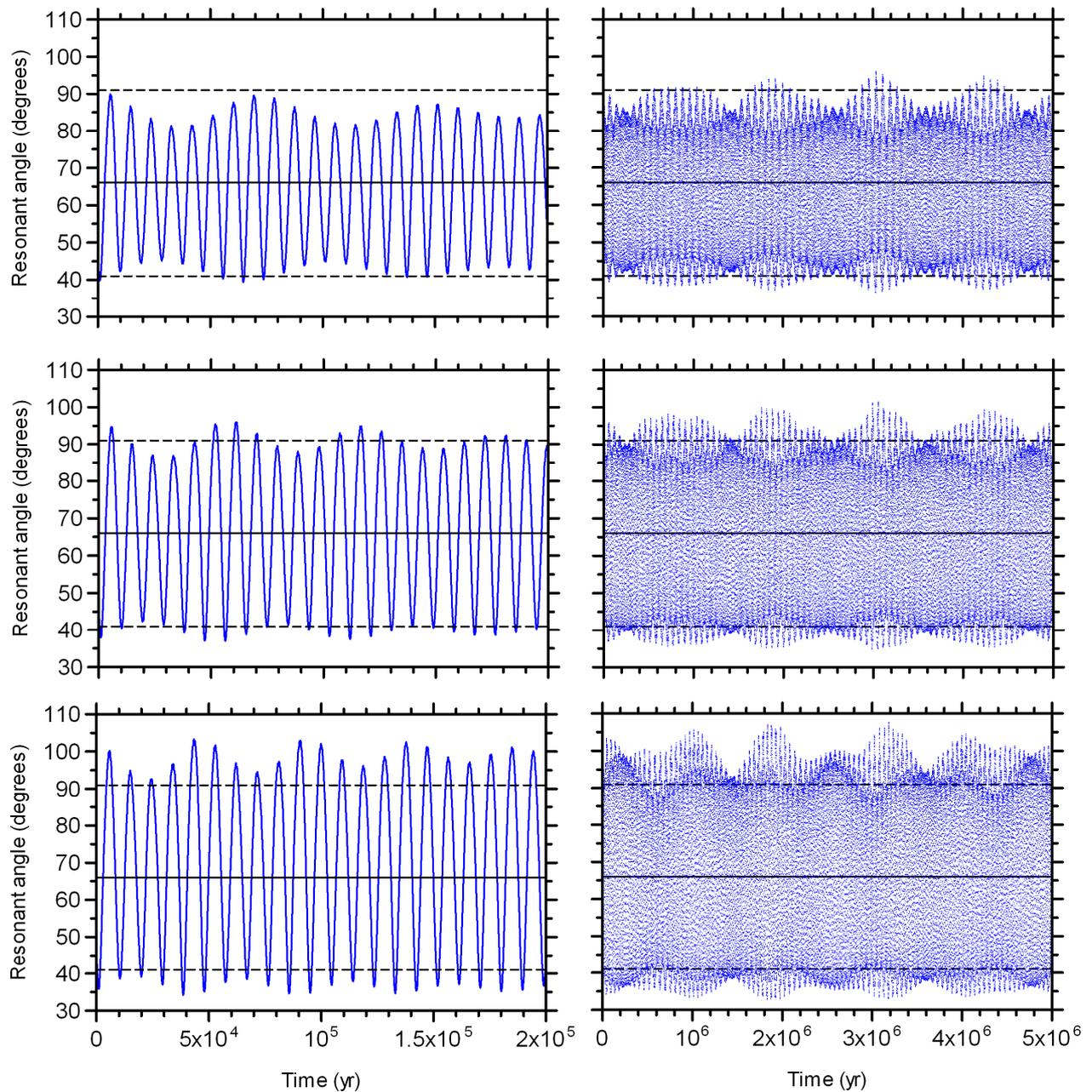

**Figure 2:** A small snapshot of the libration behaviour of three clones of 2001 QR322. The behaviour of the clone placed on the nominal orbit of the object is shown in the middle two panels, while the top and bottom show the behaviour of the two most extreme clones. Panels on the left hand side show a short term snapshot (200 kyr) of the behaviour of the objects, revealing the short term movement of the clone, in resonant angle (the distance of the object from Neptune, in its orbit, measured in degrees). The right hand panels show the behaviour over a longer timescale (5 Myr), with the data plotted as points, allowing the reader to easily see the regions in which the object spends the most time. In each plot, the solid central line marks the location of the centre of libration calculated for the entire sample ($C_L$), while the two dashed lines show the average libration amplitude for the whole sample. It is interesting to note that the clone on the nominal orbit of 2001 QR322 has the great majority of its libration maximum around the location of the average maximum amplitude for the overall sample, while the two extreme examples experience a smaller, or greater, degree of libration.

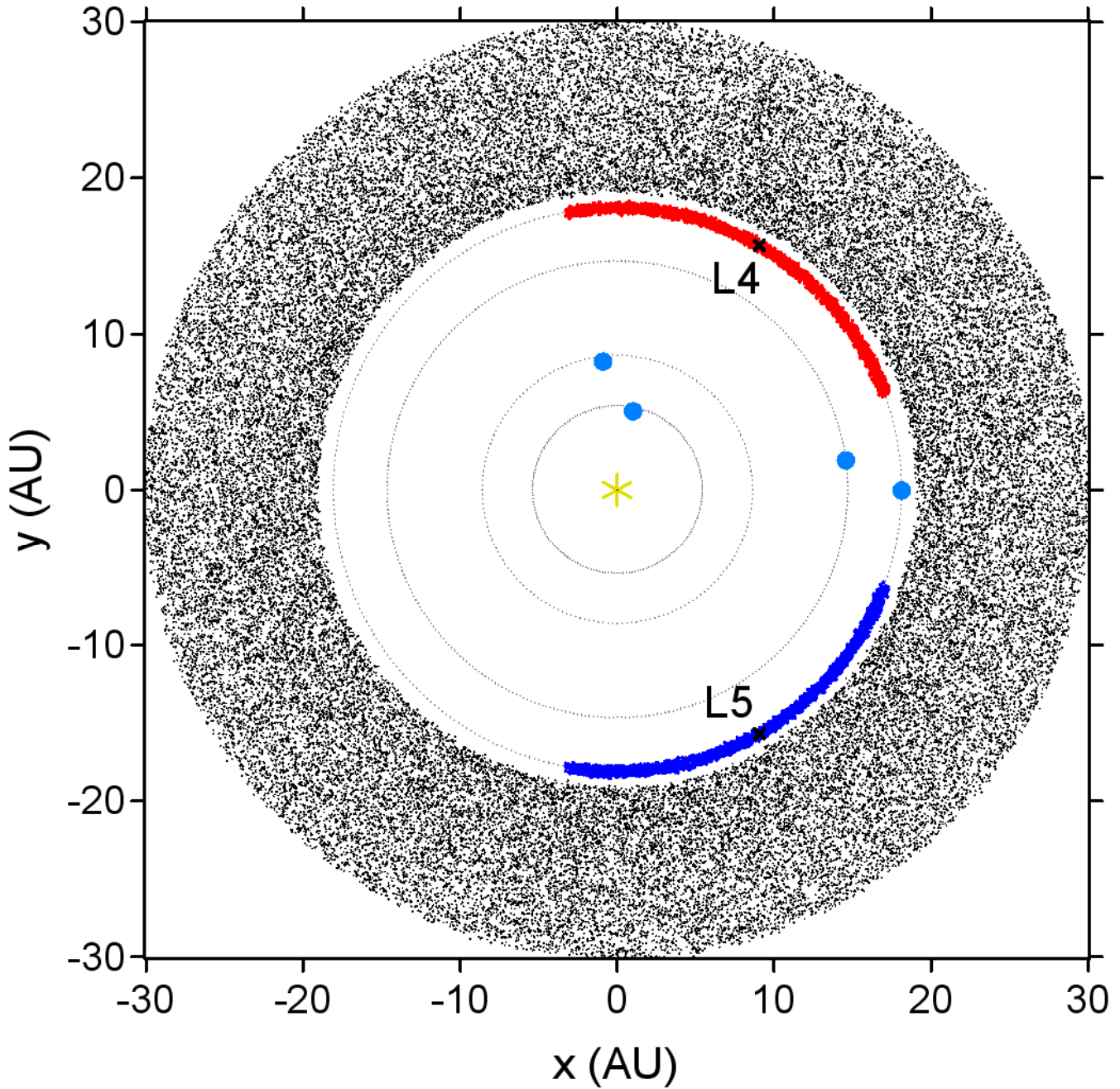

**Figure 3:** Plot showing the initial locations of all particles used to simulate the evolution of the Trojan population during Neptune's migration. Though the initial setup for each variant considered would look the same, the particles plotted here are from the scenario which involved Neptune starting at a distance of 18.1 AU from the Sun. Objects representing the pre-formed Trojans around the L4 Lagrange point are marked in red, those around the L5 point in blue, while the objects in the trans-Neptunian disk are shown in black. The pre-formed Trojans were placed with an initial displacement of up to 40º from their Lagrange point, while the disk of objects was distributed on orbits from 1 AU beyond the orbit of Neptune to 30 AU from the Sun. All particles considered were placed on initially dynamically cold orbits, with $e \sim i < 0.01$.

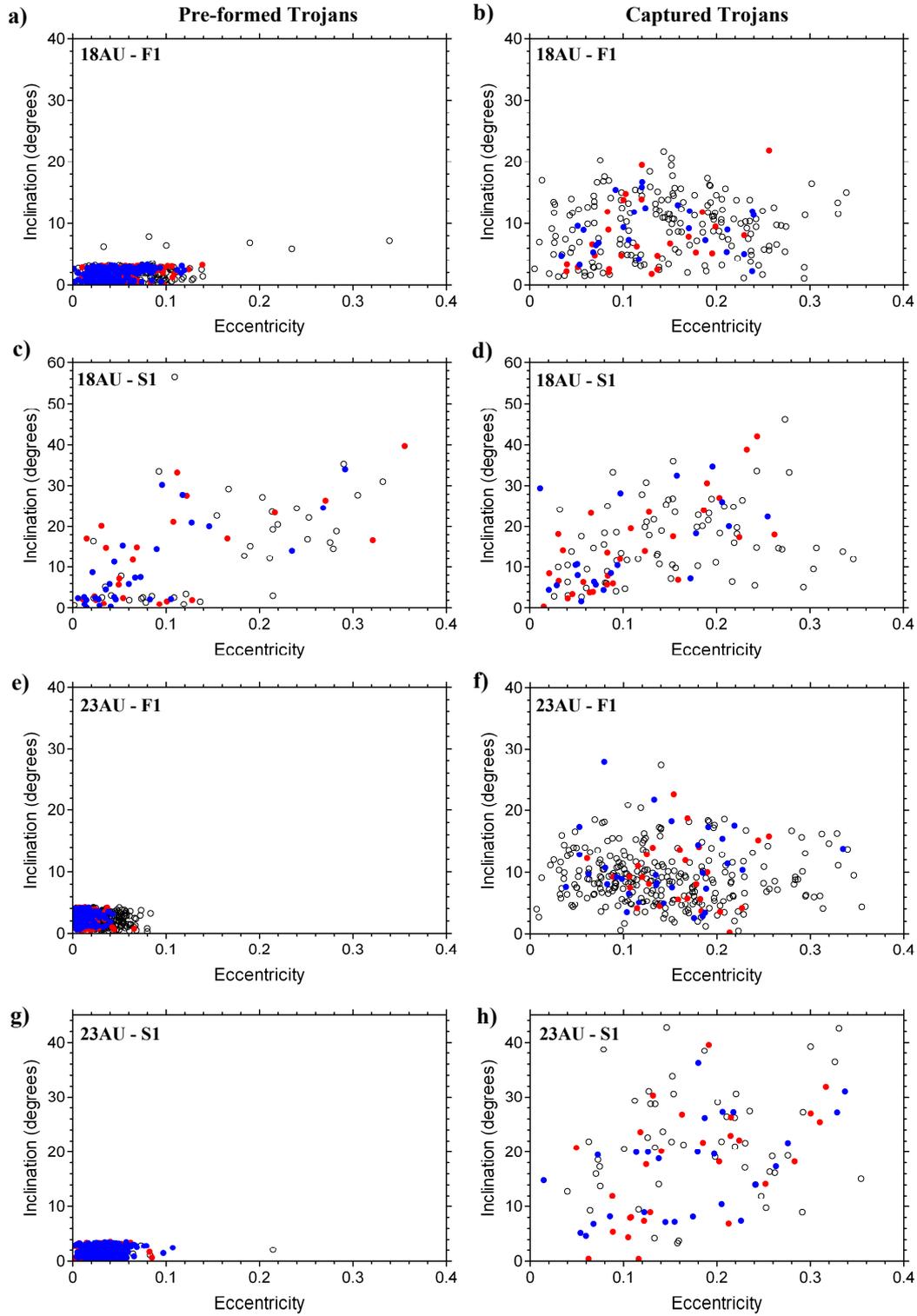

**Figure 4:** Plot showing the surviving Trojans once the giant planets had reached their current locations and stopped migration. The eight frames show the survivors for each of the four variants, with the left hand plots detailing the surviving objects from the pre-formed Trojan cloud, and the right hand plots showing those which were captured from the trans-Neptunian disk. From top to bottom, the four rows show the cases of fast migration from 18 AU (a+b), slow migration from 18 AU (c+d), fast migration from 23 AU (e+f) and slow migration from 23 AU (g+h). The objects shown in each plot correspond to the data given in Table 4 for the four runs. Objects plotted in black are located in horseshoe orbits at the end of the simulations, while those in red and blue are moving on tadpole orbits around the L4 and L5 points, respectively.

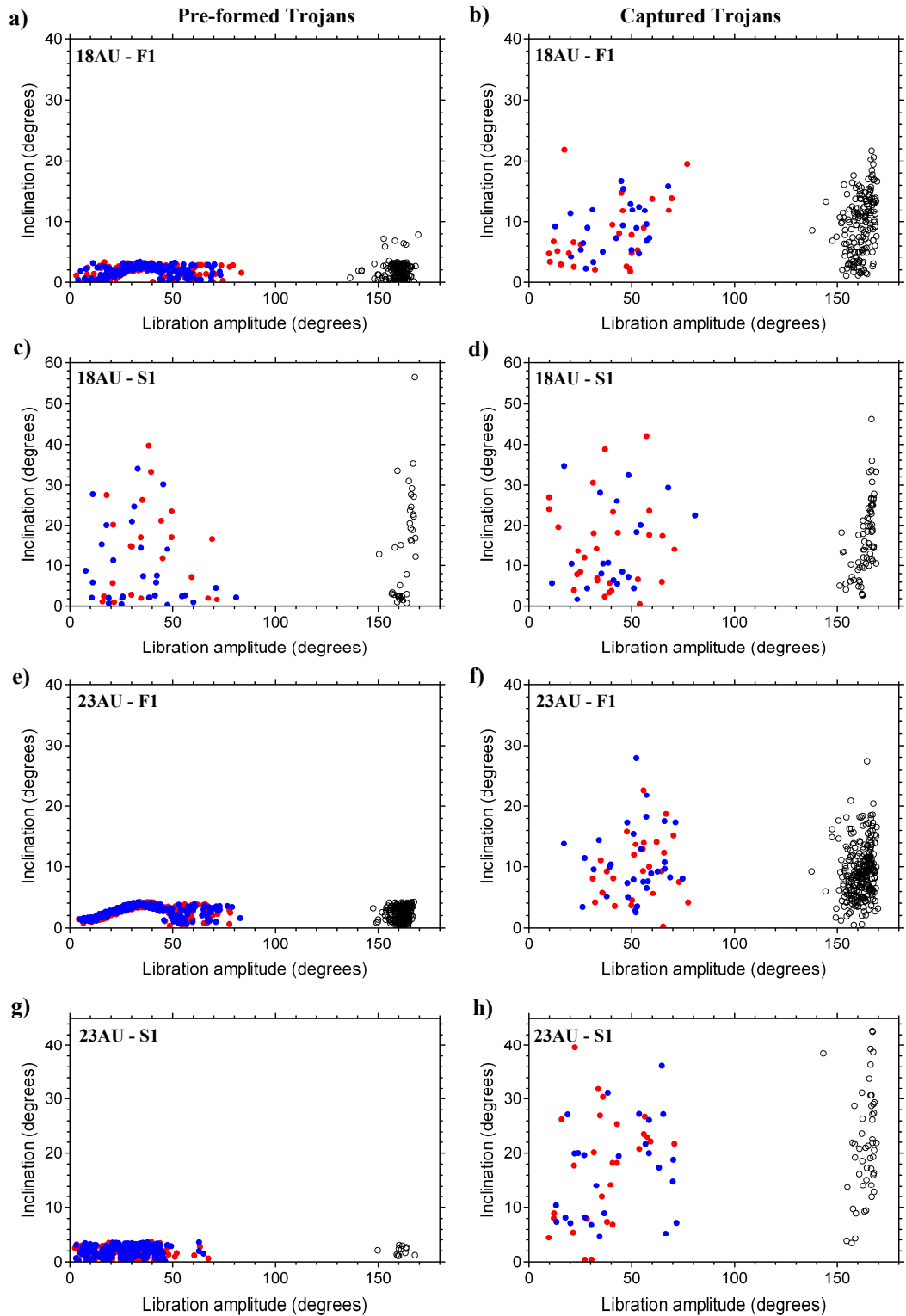

**Figure 5:** Plot showing the libration amplitudes for the Trojan objects at the end of the simulations detailed in Table 4. The distribution of plots is the same as for Figure 4 (i.e. pre-formed objects in the left hand plots, captured objects on the right; rows 1-4 show the cases 18AU-F1, 18AU-S1, 23AU-F1 and 23AU-S1, respectively; objects in black finish the simulation on horseshoe-type orbits, while those in red (L4) and blue (L5) are located in tadpole orbits).

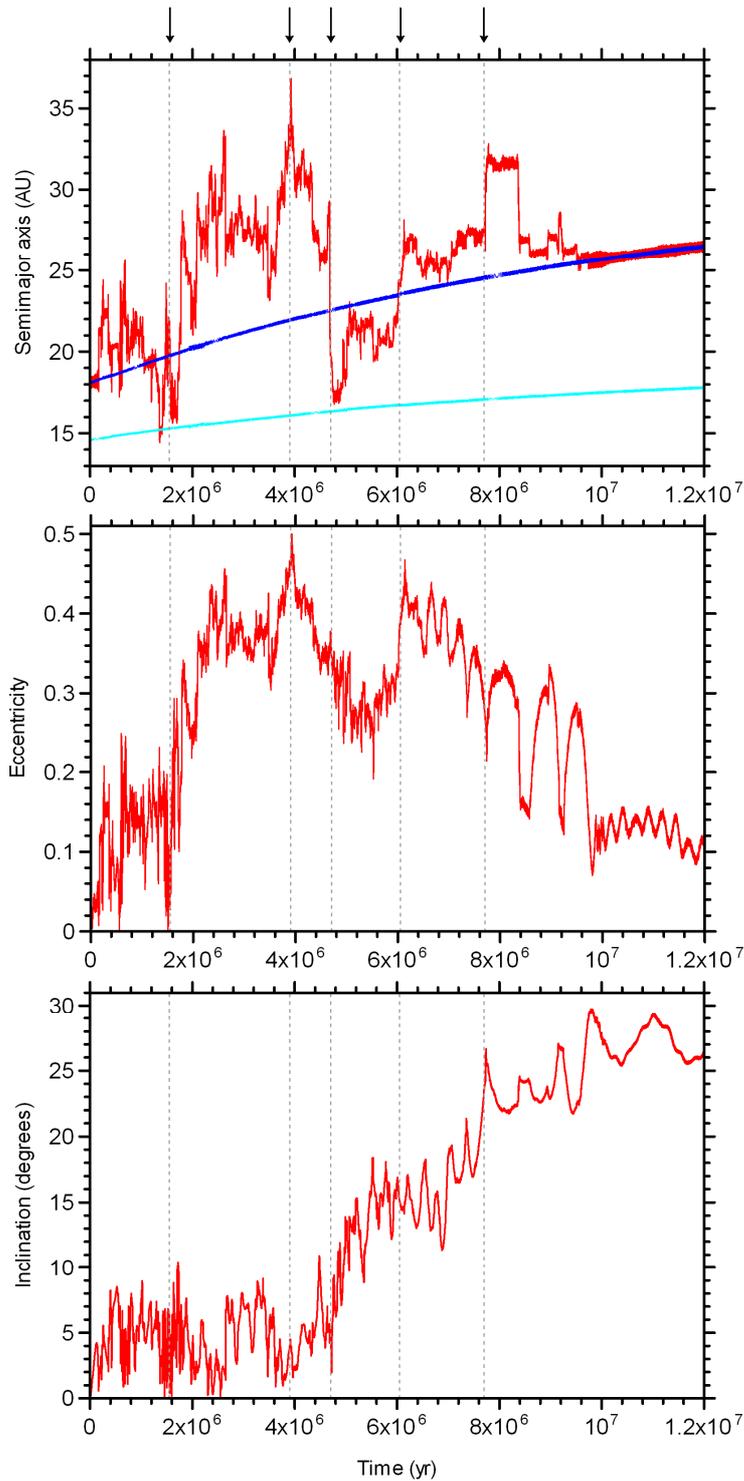

**Figure 6:** An example of a pre-formed Trojan being first lost, then recaptured, from the Trojan cloud during the slow migration of Neptune from 18.1 AU. The plots detail the first 12 Myr of the objects evolution, with its semi-major axis, eccentricity and inclination being plotted at top, middle and bottom, respectively. The evolution of Uranus' and Neptune's semi-major axes is shown in the upper panel (cyan and blue curves, respectively). The majority of the clone's evolution is spent drifting within the Trojan cloud, and migrating along with Neptune (this behaviour continues unchanged until the end of the simulations, at 50 Myr; $\tau = 10$ Myr). The object experiences a number of close encounters with Uranus and Neptune (marked by sudden large changes in the orbital elements; a few such encounters are marked on the plot (vertical dashed lines)), together with a number of short term resonant captures (periods of stable, albeit oscillating, orbital elements – e.g. at 7.6 Myr). Finally, the object is recaptured to the Trojan family, albeit with greatly excited inclination.